\documentclass[5p]{elsarticle}
\usepackage{graphicx}
\usepackage{amsmath}
\usepackage{amssymb}
\usepackage{subfigure}
\usepackage{psfrag}
\usepackage{color}

\begin{document}

\title{Radiation of scalar oscillons in 2 and 3 dimensions}


\author[rmki]{Gyula Fodor}
\author[rmki,tours]{P\'eter Forg\'acs}
\author[elte]{Zal\'an Horv\'ath}
\author[elte]{M\'ark Mezei}
\address[rmki]{MTA RMKI, H-1525 Budapest 114, P.O.Box 49, Hungary}
\address[elte]{Institute for Theoretical Physics, E\"otv\"os University,
H-1117 Budapest, P\'azm\'any P\'eter s\'et\'any 1/A, Hungary}
\address[tours]{LMPT, CNRS-UMR 6083, Universit\'e de Tours,
Parc de Grandmont, 37200 Tours, France}


\begin{abstract}
  The radiation loss of small-amplitude radially symmetric oscillons
  (long-living, spatially localized, time-dependent solutions) in two-
  and three-dimensional scalar field theories is computed analytically
  in the small-amplitude expansion.  The amplitude of the radiation is
  beyond all orders in perturbation theory and it is determined using
  matched asymptotic series expansions and Borel summation.  The
  general results are illustrated on the case of the two- and
  three-dimensional sine-Gordon theory and a two-dimensional $\phi^6$
  model.  The analytic predictions are found to be in good agreement
  with the results of numerical simulations of oscillons.
\end{abstract}

\maketitle

\section{Introduction}\label{s:intro}

There is an increasing interest in long-living, spatially localized
classical solutions in field theories -- {\sl oscillons} -- exhibiting
nearly periodic oscillations in time
\cite{dashen}--\cite{sicilia}. For the two and more spatial
dimensional sine-Gordon theory, following the original naming in
\cite{BogMak2}, these objects are generally called pulsons
\cite{Chris}--\cite{Bratsos}. Oscillons resemble the ``true'' (time
periodic and of finite energy) breathers of the one-dimensional
sine-Gordon (SG) theory, but unlike true breathers they are
continuously losing energy by radiating slowly. Just like a breather,
an oscillon possesses a localized ``core'', but it also has a
``radiative'' region outside of the core. Oscillons appear 
from rather generic initial data in the course of time evolution,
in an impressive number of physically
relevant theories including the bosonic sector of the standard model
\cite{Farhi05}--\cite{Borsanyi2}. Moreover they form in physical
processes making them of considerable importance
\cite{Kolb:1993hw}--\cite{GleiserTHor08}.

A crucial physical characteristic of oscillons is the amplitude of the
outgoing wave determining their lifetime. In a previous work we have
been able to perform an analytic calculation of the radiation
amplitude of oscillons in scalar theories in the limit of small
oscillon amplitudes in one spatial dimension \cite{FFHM}. In this
limit one can perform an expansion yielding breather-like
configurations with spatially localized cores. The small-amplitude
expansion yields an asymptotic series for the core, but misses a
standing wave tail whose amplitude is exponentially small with respect
to that of the core \cite{SK}. Therefore in order to compute the
radiative amplitude, more precisely its leading part, implies to go
beyond all orders in perturbation theory. In the present paper
building on the results of \cite{FFHM} we develop a method to compute
the radiation amplitude in $D=2$ and $D=3$ spatial dimensions in the
small-amplitude expansion. The
main result of this paper can be summarized by the following simple
formula determining the energy loss of a small-amplitude oscillon in
$D$ spatial dimensions ($D<4$):
\begin{equation}
\label{e:genradlaw}
\frac{\mathrm{d} E}{\mathrm{d} t}=- \frac{A_{D}}{\varepsilon^{D-1}}
\exp\left(-\frac{B_{D}}{\varepsilon}\right)\,,
\end{equation}
where $A_D\,,B_D$ are given in Eq. \eqref{e:symradlaw} as functions of $D$ and
of the coefficients of the scalar potential.
Although our analytic calculations are carried out in the limit of the oscillon
amplitude going to zero, the results are valid for non infinitesimal
values of the amplitude. We have found satisfactory agreement between the predicted
energy loss Eq. \eqref{e:genradlaw} and the ``measured'' one (by numerical simulations)
for the SG theory in $D=2$ and $D=3$ and in a $\phi^6$-type model investigated in
Ref.\ \cite{FFHM} in $D=2$.
Our results also imply that no breathers
depending on continuous parameters can exist for $D>1$ for arbitrary
scalar potentials. Note that in $D=1$ the SG theory is
the {\sl only} one with analytic potential admitting a breather family \cite{Kichenassamy}.

\section{Small-amplitude expansion}

We consider spherically symmetric solutions of a real scalar theory in
a $D+1$ dimensional Minkowski space-time, with a self-interaction
potential $U(\phi)$. The equation of motion is
\begin{equation}\label{e:evol}
 -\frac{{\partial}^2\phi}{{\partial} t^2}+
\frac{{\partial}^2\phi}{{\partial} r^2}
+\frac{D-1}{r} \,\frac{{\partial}\phi}{{\partial} r}
= U'(\phi)=\phi +\sum\limits_{k=2}^{\infty}g_k\phi^k\,.
\end{equation}
The mass of the field is chosen to be $1$, and the derivative of the
potential $U(\phi)$ is expanded as a power series in $\phi$, where the
$g_k$ are constants. In the following unless otherwise stated we shall
consider potentials which are symmetric around their minima,
i.e.\ $g_{2k}=0$.

The small-amplitude expansion of solutions of Eq. \eqref{e:evol}
can be written as
\begin{equation}
\phi=\sum_{k=1}^\infty\varepsilon^k \phi_k \,, \label{e:sumphi}
\end{equation}
where $\varepsilon\ll 1$ is the expansion parameter .
Demanding the functions $\phi_k$ be bounded they all turn out to
be periodic in time.
For the class of symmetric potentials $\phi_{2k}=0$, $k=1\,,2,\ldots$,
and the first two non-vanishing amplitudes are given as
\begin{eqnarray}
\phi_1&=&p_1\cos (\omega t)\,,\label{e:phi1}\\
\phi_3&=&p_3\cos (\omega t)-\frac{\lambda}{24}p_1^3\cos(3\omega t)
\label{e:phi3}\,,
\end{eqnarray}
where $\lambda=-3g_3/4$, the frequency,
$\omega=\sqrt{1-\varepsilon^2}$, and the functions $p_1$ and $p_3$
depend only on the radial coordinate $r$.  Introducing $S$ and $Z$ by
\begin{equation}\label{e:SZdef}
p_1=\frac{S}{\sqrt{\lambda}}\,, \quad
p_3=\mu Z\,, \quad
\mu=\frac{1}{\lambda^2\sqrt{\lambda}}
\left(\frac{1}{24}\lambda^2+\frac{5}{8}g_5\right) ,
\end{equation}
the constants $g_k$ get eliminated from the equations determining
$p_1$ and $p_3$,
\begin{eqnarray}
&&\frac{{\rm d}^2S}{{\rm d}\rho^2}
+\frac{D-1}{\rho}\,\frac{{\rm d}S}{{\rm d}\rho}
-S+S^3=0\,,\label{e:eqS}\\
&&\frac{{\rm d}^2Z}{{\rm d}\rho^2}
+\frac{D-1}{\rho}\,\frac{{\rm d}Z}{{\rm d}\rho}\,
-Z+3S^2Z-S^5=0 \,,\label{e:eqZ}
\end{eqnarray}
where a rescaled radial coordinate has been introduced through
$\rho=\varepsilon r$.
Eq.\ \eqref{e:eqS} admits regular solutions decaying exponentially
in dimensions $D<4$, implying similar properties for the regular
solutions of Eq.\ \eqref{e:eqZ}. From now on we will only discuss the fundamental
solution of Eq.\ \eqref{e:eqS} without nodes. 
On solutions with nodes see Ref. \cite{FFHL2}.
The amplitudes $\phi_{2k+1}$, $k\geq2$, are determined
by linear inhomogeneous equations analogous to \eqref{e:eqZ}, and they
all decay exponentially.
The small amplitude series \eqref{e:sumphi} does not converge in general -- it is
an {\sl asymptotic} one.
This is closely related to the fact that no true breathers exist in generic theories
because of the radiative degrees of freedom.
Nevertheless the asymptotic series \eqref{e:sumphi} yields an excellent
approximation for the spatially localized ``core'' part of the oscillon
(for details see Ref. \cite{FFHL2}).

It turns out that the radiative amplitude of an oscillon is of the
order of $\exp\left(-B_{D}/\varepsilon\right)$ i.e.\ beyond reach of
finite order perturbation theory in $\varepsilon$. As found in Ref.\
\cite{FFGR} any slowly radiating oscillon solution can be well
approximated by a suitable exactly time periodic, ``quasibreather''
(QB) having a well-localized core and an inevitable standing wave
tail, whose amplitude is in some sense minimal.  In Section 4.\ the
amplitude of this standing wave tail and that of the outgoing
radiation will be directly related.  To find this amplitude
using the Segur-Kruskal method \cite{SK}, we have to analytically continue
our functions to the complex plane and find the singularities closest
to the real axis as a first step. Because (\ref{e:eqZ}) is linear in
$Z$ its solution has a singularity exactly where $S$ has. Thus, we
have to find the singularities of the master equation (\ref{e:eqS}). In
the neighbourhood of the closest singularity to the real axis we compute
the nonperturbative correction,  
which determines the radiation field of the
oscillon after matching it to the solution on the real axis.
$S$
has simple pole singularities on the imaginary axis at $\rho=\pm
i\,P_{D}$. Let us measure the distances from the upper singularity as
\begin{equation}
\rho=i\,P_{D}+R\,.	
\end{equation}
Then the solution of (\ref{e:eqS}) for small $R$ can be written in the
following series form
\begin{equation}\label{e:srser}
S=-\frac{i\,\sqrt2}{R}+\frac{\sqrt2\,(D-1)}{6P_{D}} +\mathcal{O}(R) \,.
\end{equation}
We note that for $D>1$ the series (\ref{e:srser}) will
also contain non-analytic terms, the lowest order being $R^3\ln R$.
The position of the pole can be determined both by Pad\'e's approximation and by
integrating the master equation numerically along the imaginary
axis. Table \ref{t:pole} gives the results of these numerical
approaches.
\begin{table}[!htbp]
\begin{center}
\begin{tabular}{|c|c|}
\hline
$D$  & $P_{D}$ \\
\hline
1 & 1.57080\\
2 & 1.09256\\
3 & 0.60218\\
\hline								
\end{tabular}
\end{center}
\caption{The distance between the real axis and the pole of the fundamental
  solution of the master equation (\ref{e:eqS}) as a function of
  the number of spatial dimensions $D$. \label{t:pole}}
\end{table}
For $D=4$ $P_{D}=0$, i.e.\ the pole is in the origin.
It has been shown in Ref.\ \cite{FFHL2} that
there are no regular and localized solutions of Eq.\ \eqref{e:eqS} for $D\geq4$.
This implies that in dimensions $D\geq4$ the oscillon core is not localized.
As it will be shown the pole term $P_D$ appears in the exponential in the radiation law,
and from Table \ref{t:pole} it follows
that oscillons in higher dimensions will radiate faster.

The solution of the first inhomogenous equation (\ref{e:eqZ})
in the neighborhood of the singularity takes the form
\begin{equation}\label{e:zrser}
Z=-\frac{i\,2\sqrt2}{3}\,\frac{1}{R^3}
-\frac{8\sqrt2\,(D-1)}{15\,P_{D}}\,\frac{\ln R}{R^2}
+\frac{z_{-2}}{R^2}+ \ldots\ .
\end{equation}
The non-analytic terms represent dimensional corrections to the
one-dimensional results. The free constant $z_{-2}$ is due to the
solution of the homogeneous part of the equation for $Z$ and is
determined by the matching conditions, i.e.\ by using the fact that
$Z$ is decaying on the real axis.

We have seen that the small-amplitude expansion yields time periodic and
spatially localized (of finite energy) configurations which appear to be
breathers. On the other hand any time periodic solution of Eq.\ \eqref{e:evol}
can be expanded into a Fourier series as:
\begin{equation}\label{e:fourdec}
\phi=\sum_{n=0}^\infty\cos(n\omega t)\, \Phi_n \ .
\end{equation}
$\Phi_{2n}=0$ in the case of symmetric
potentials presently considered. For later use we compare the two different
expansions, (\ref{e:sumphi}) and \eqref{e:fourdec}:
\begin{equation}\label{e:p13}
\Phi_1=\varepsilon p_1+\varepsilon^3p_3
+\mathcal{O}(\varepsilon^5) \,, \quad
\Phi_3=-\frac{\lambda}{24}\varepsilon^3p_1^3
+\mathcal{O}(\varepsilon^5) \, .
\end{equation}

Next we determine the behaviour of the Fourier modes, $\Phi_n$, near
the singularity. We define a new spatial coordinate $y$ by
$R=\varepsilon y$, which is related to $r$ as:
\begin{equation}
r=\frac{i\,P_{D}}{\varepsilon}+ y\,. \label{e:yr}
\end{equation}
The ``inner region'' $R\ll 1$ is not small in the $y$ coordinates; if
$\varepsilon\to0$ then $\varepsilon \vert y\vert=\vert R\vert\ll 1$
but $\vert y\vert\to\infty$.  Then from Eqs.\ (\ref{e:srser}) and
(\ref{e:zrser}) the asymptotic behaviour for $\vert y\vert\to\infty$
of $S$ and $Z$ can be rewritten as
\begin{align}
\varepsilon S=&\frac{-i\sqrt2}{y}
+\varepsilon\frac{\sqrt2(D-1)}{6P_{D}} +\ldots\\
\varepsilon^3 Z=&-\frac{i2\sqrt2}{3}\,\frac{1}{y^3}
-\varepsilon \ln \varepsilon\,\frac{8\sqrt2(D-1)}{15P_{D}}
\,\frac{1}{y^2}\nonumber\\
&-\varepsilon\,\frac{8\sqrt2\,(D-1)}{15\,P_{D}}\,
\frac{\ln y}{y^2}+ \varepsilon \frac{z_{-2}}{y^2}+ \ldots \,,
\label{e:zrser2}
\end{align}
and from Eqs.\ (\ref{e:p13}), (\ref{e:SZdef}) we obtain
\begin{align}
&\Phi_1=-\frac{i \sqrt2}{\sqrt\lambda}\,\frac{1}{y}
-\frac{2i\mu\sqrt2}{3}\,
\frac{1}{y^3}-\varepsilon \ln \varepsilon\,
\frac{8\mu\sqrt2(D-1)}{15\,P_{D}}\,\frac{1}{y^2}\nonumber\\
&+\varepsilon\Biggl[\frac{\sqrt2(D-1)}{6\sqrt\lambda P_{D}}
+\mu\left(\frac{z_{-2}}{y^2}-\frac{8\sqrt2\,(D-1)}{15\,P_{D}}
\,\frac{\ln y}{y^2}
\right)\Biggr]+\ldots \,,\label{e:bphi1}\\
&\Phi_3=-\frac{i \sqrt2}{12\sqrt\lambda}\,\frac{1}{y^3}
+\varepsilon\,\frac{\sqrt2(D-1)}{24\sqrt\lambda\,P_{D}}\,
\frac{1}{y^2}+\ldots \ .\label{e:bphi3}
\end{align}
We shall impose Eqs.\ \eqref{e:bphi1}, \eqref{e:bphi3} as boundary conditions
to construct the inner solution (close to the singularity).
We note that there is also a term
proportional to $\varepsilon \ln \varepsilon/y^4$ in $\Phi_3$ arising
from $\varepsilon^5$ order terms in the small-amplitude
expansion.

To leading-order, i.e.~setting $\varepsilon=0$ in \eqref{e:bphi1} and
\eqref{e:bphi3}, we recover the one-dimensional result obtained in
\cite{FFHM}. The first dimensional correction comes from the
$\varepsilon \ln \varepsilon$ terms. As we shall see, this will give a
multiplicative correction factor to the energy loss rate. In the case
of the SG model for $D=1$ the breathers do not radiate. 
To determine the radiation loss in the 
SG models for $D>1$ we have to calculate the next order
corrections, arising from the $\varepsilon$ terms.

\section{Fourier mode expansion near the pole}

Let us now directly Fourier decompose our field equation
(\ref{e:evol}) using (\ref{e:fourdec}), without employing the
small-amplitude expansion,
\begin{equation}\label{e:modes}
\left[\frac{{\rm d}^2}{{\rm d} r^2}
+ \frac{D-1}{r}\,\frac{{\rm d}}{{\rm d}r}
+(n^2\omega^2 -1)\right] \Phi_n  =F_n \,,
\end{equation}
where the nonlinear terms on the right hand side have the form
\begin{equation}
F_n=\frac{g_3}{4}\,\sum_{m,p,q=1}^{\infty}\Phi_m\Phi_p\Phi_q\,
\delta_{n,\,\pm m \pm p\pm q}+\ldots \ .
\end{equation}
Then in the inner region the mode equations \eqref{e:modes} keeping
also $\cal{O}(\varepsilon)$ correction can be written as
\begin{equation}\label{e:modeseq}
\left[\frac{{\rm d}^2}{{\rm d} y^2}
+ \varepsilon\frac{D-1}{i\,P_{D}}\,\frac{{\rm d}}{{\rm d}y}
+(n^2 -1)+{\cal{O}}(\varepsilon^2)\right] \Phi_n  =F_n \ .
\end{equation}

By expanding equations \eqref{e:modeseq} directly in powers of
$1/y$, including $\ln\,y$ terms when necessary one can obtain
Eqs.\ \eqref{e:bphi1}, \eqref{e:bphi3}.
Technically it is easier to obtain \eqref{e:bphi1} and \eqref{e:bphi3}
this way then by the small $\varepsilon$ expansion.

Imposing Eqs.\ \eqref{e:bphi1}, \eqref{e:bphi3} (and the corresponding ones for $n>3$)
as boundary conditions for $\mathrm{Re}\,y\to\infty$
on the Fourier modes $\Phi_n$ defines a unique solution of the system \eqref{e:modeseq}.
This corresponds to the behavior of
the (real) solution for which all $\Phi_n$ decay exponentially when
$r\to\infty$ on the real axis.
Since in general no true breather exists, this decaying solution
is singular at $r=0$.
Its extension to the complex plane is not real on the $\mathrm{Re}\,y=0$ axis,
the imaginary parts of the modes $\Phi_n$ satisfy homogeneous equations
to leading-order in $1/y$. Specifically, the solution for $\Phi_3$ behaves as
\begin{equation}
\mathrm{Im}\,\Phi_3=\nu_3\exp(-i\sqrt{8}\, y) \ \ \text{for} \ \
 \mathrm{Re}\,y=0 \,. \label{e:imphi3}
\end{equation}
The constant $\nu_3$ determines the leading-order part of the radiation
amplitude in the $\Phi_3$ mode which we aim to compute.
A method to find the value of $\nu_3$ developed in Ref.\ \cite{SK}
is to integrate Eqs.\ \eqref{e:modeseq}
along a constant $\mathrm{Im}\,y$ line numerically,
starting from the values given by \eqref{e:bphi1}, \eqref{e:bphi3} for large $\vert y\vert$.
An analytic method, using Borel summation, has been proposed
in Refs.\ \cite{Hakim,Pomeau} and has
been adapted for one-dimensional scalar oscillons in \cite{FFHM}.

Since to zeroth order in $\varepsilon$ Eq.\ \eqref{e:modeseq} is the same
as for $D=1$, the constant $\nu_3$ is independent of
$D$ for $\varepsilon\to0$, corresponding to near
threshold states with $\omega\to1$.
In order to find the corrections of order
$\cal{O}(\varepsilon \ln \varepsilon)$ and
of $\cal{O}(\varepsilon)$
to the value of $\nu_3$, it is sufficient to
solve the mode equations linearized about the one-dimensional
solution. Denoting the solution of the equations for $D=1$ by
$\Phi_n^{D=1}+{\cal{O}}(\varepsilon^2)$,
and writing $\Phi_n=\Phi_n^{D=1}+\widetilde{\Phi}_n$,
the equations linearized in $\varepsilon$ around $D=1$ take the form
\begin{equation}\begin{split}
&\left[\frac{{\rm d}^2}{{\rm d} y^2} +(n^2 -1)\right]
\widetilde{\Phi}_n +\varepsilon\frac{D-1}{i\,P_{D}}\,
\frac{{\rm d}}{{\rm d}y}\,\Phi_n^{D=1}\\
&\qquad\qquad=\sum_{m=odd}\left.\frac{\partial F_n}{\partial \Phi_m}
\right|_{\Phi_n=\Phi_n^{D=1}}\,\widetilde{\Phi}_m\,. \label{e:inhom}
\end{split}\end{equation}
$\widetilde{\Phi}_n$ contains in general corrections
of order $\cal{O}(\varepsilon \ln \varepsilon)$ and
of $\cal{O}(\varepsilon)$.
The solution of the linearized equations to order $\cal{O}(\varepsilon \ln \varepsilon)$
can be explicitly written as
\begin{equation}
\widetilde{\Phi}_n=\varepsilon \ln \varepsilon\,
C\frac{{\rm d}}{{\rm d}y}\,\Phi_n^{D=1}\,, \label{e:zeromode}
\end{equation}
where $C$ is an arbitrary constant independent of $n$. Eq.\ \eqref{e:zeromode}
corresponds simply the zero mode when $D=1$ in Eq. \eqref{e:inhom}. The
constant $C$ is determined by the matching conditions \eqref{e:bphi1}
and \eqref{e:bphi3}:
\begin{equation}
C=i\sqrt{\lambda}\mu\,\frac{8\,(D-1)}{15\,P_{D}}\,. \label{e:eqC}
\end{equation}
In order to apply the Borel summation method described in detail in \cite{FFHM},
one expands the mode equations near the singularity, e.g.\ for $\Phi_3^{D=1}$:
\begin{equation}
\Phi_3^{D=1}=i\,\sum_{k=2}^{\infty} \frac{B_k}{y^{2k-1}}\,, \label{e:phi3d1}
\end{equation}
and determines the growth of the coefficients $B_k$ for large $k$ as a first step.
As found in Ref.\ \cite{FFHM}:
\begin{equation}
B_k\sim K^{D=1}\,(-1)^k\,\frac{(2k-2)!}{8^{k-1/2}} \,.
\end{equation}
The Borel sum of Eq.\ \eqref{e:phi3d1} yields the imaginary part
(in the case of symmetric potentials) of $\Phi_3^{D=1}$ and by comparing it
to Eq.\ \eqref{e:imphi3} we have found
the connection between the constant $K$ and $\nu_3$, yielding
$\nu_3^{D=1}=K^{D=1}\,\pi/2$. Expanding the mode equation near the
singularity and calculating $K$ from the behaviour of $B_k$ for large
$k$ is a much easier and more precise way to calculate $\nu_3$ than
numerical integration. Since the leading-order
dimensional corrections, of order $\varepsilon \ln \varepsilon$ are
given by \eqref{e:zeromode}, the dominant dimensional contribution
to the value of $K$ is given as
\begin{equation}\label{e:khidim}
K=K^{D=1}\,\left(1-\varepsilon \ln \varepsilon\,i\,\sqrt8 C \right)\,,
\end{equation}
where $C$ is given in \eqref{e:eqC}.
The relation $\nu_3=K\pi/2$ still remains true. In
general, this expression already gives satisfactory results for the
energy loss of oscillons, except for the SG theory, where $K^{D=1}=0$. Then the
correction of order $\cal{O}(\varepsilon)$ become essential.

We now outline the computations of the dimensional corrections of $\cal{O}(\varepsilon)$
to the SG theory, in which case this correction is the leading one to the
radiation amplitude. When $K^{D=1}\neq 0$ there are other contributions to this order
besides the dimensional correction, such as
$1/y$ corrections to Eq.\ \eqref{e:imphi3} (see Eq. (23) of \cite{FFHM}) and
the interaction of the outgoing wave with the core and these we do not
compute here.
It has already been pointed out that to each term of
order $\cal{O}(\varepsilon \ln \varepsilon)$ there corresponds a term of order
$\cal{O}(\varepsilon)$, which is obtained by changing $\ln \varepsilon$ to $\ln y$.
Hence introducing $\overline{\Phi}_n$ by
\begin{equation}
\widetilde{\Phi}_n=\varepsilon \ln \varepsilon\,C
\frac{{\rm d}}{{\rm d}y}\,\Phi_n^{D=1}+\varepsilon \,
\left(C\ln y\,\frac{{\rm d}}{{\rm d}y}\,\Phi_n^{D=1}\,
+\overline{\Phi}_n\right)\,,\label{e:dimcor}
\end{equation}
the linearized equation \eqref{e:inhom} takes the form
\begin{align}
&\left[\frac{{\rm d}^2}{{\rm d} y^2} +(n^2 -1)\right] \,
\overline{\Phi}_n + C \left(\frac{2}{y}\,
\frac{{\rm d^2}}{{\rm d}y^2}\,\Phi_n^{D=1}-\frac{1}{y^2}\,
\frac{{\rm d}}{{\rm d}y}\,\Phi_n^{D=1}\right)\nonumber\\
&\ +\frac{D-1}{i\,P_{D}}\,\frac{{\rm d}}{{\rm d}y}\,\Phi_n^{D=1}
=\sum_{m=odd}\left.\frac{\partial F_n}{\partial \Phi_m}
\right|_{\Phi_m=\Phi_m^{D=1}}\,\overline{\Phi}_m\,. \label{e:epscor}
\end{align}
The advantage of using $\overline{\Phi}_n$ is that there are no $\ln y$
terms in its $1/y$ expansion.
Expanding now $\overline{\Phi}_3$ as
\begin{equation}
\overline{\Phi}_3=\sum_{k=1}^{\infty} \frac{\beta_k}{y^{2k}} \,,
\end{equation}
from \eqref{e:epscor} it follows that to leading-order
\begin{equation}
 (2k-2)(2k-1)\beta_{k-1}+8\beta_{k}-(2k-1)\,\frac{D-1}{P_{D}}\,B_k=0\,.
\label{e:beta}
\end{equation}
To find the large $k$ behaviour of $\beta_k$, we consider the following Ansatz:
\begin{equation}
\beta_k \sim L\,(-1)^k\,\frac{(2k)!}{8^{k}}
+M\,(-1)^k\frac{(2k-1)!}{8^{k}} \,,
\end{equation}
where $L$ and $M$ are the constants to be determined.
Eq.~\eqref{e:beta} yields $L$,
\begin{equation}
L=\frac{\sqrt2\, (D-1)}{8P_{D}}\,K^{D=1}
\end{equation}
while $M$ is still arbitrary to this order in $k$.  For the
SG model $K^{D=1}=0$, and by substituting the $1/y$ expansion
into Eq. \eqref{e:epscor} we obtain a higher-order
asymptotic formula for the behaviour of $\beta_k$ for large values of $k$,
\begin{equation}
\beta_k\sim M(-1)^k\,\frac{(2k-1)!}{8^{k}}\;
\left(1+\frac{1}{k}+\frac{3}{4k^2}\right)\,.
\end{equation}
The value of $M$ can then be obtained to a good precision by
explicitly calculating the coefficients $\beta_k$ in the expansion
of \eqref{e:epscor} to moderately high orders. Although we always
work with a truncated SG potential and with a finite number of modes
the value of $K^{D=1}$ can be made very small by using $\Phi_1-\Phi_9$
and truncating the potential at $12^{th}$ order. For the constant
$K$ of the SG theory we obtain
\begin{equation}\label{e:ksg}
K=\varepsilon M \ , \quad
M=0.6011\frac{D-1}{P_{D}} \,.
\end{equation}
We note that although a term proportional to $z_{-2}$ (defined in
\eqref{e:zrser}) also appears in the obtained values of $\beta_k$, its
coefficient quickly tends to zero when increasing the order of truncation.
Repeating the Borel summation argument for this case, it turns
out that $\nu_3=K\pi/2$ still holds, giving the behaviour of
$\mathrm{Im}\,\Phi_3$ on the imaginary axis in the neighbourhood of the
singularity. Naturally, the next step is to continue the imaginary
correction back to the real axis.

\section{Continuation to the real axis}

The standing wave tail of a small amplitude time periodic QB 
satisfies to leading order a homogeneous linear equation given by the left hand side of
\eqref{e:modes}. Since the size of a QB grows
proportionally to $1/\varepsilon$, it is possible to consider an
inner region, which is, however, outside of the domain around $r=0$
where the asymptotically decaying modes, $\Phi_n$, get large. In the previous section we constructed the function $\delta\Phi_3$ by Borel-summing the asymptotic series. We extend $\delta\Phi_3$ to
the inner region assuming that it behaves as \eqref{e:imphi3} close to the upper pole,
and as $\mathrm{Im}\,\Phi_3=-\nu_3\exp(i\sqrt{8}\, y)$ near the lower
pole, where $r=-iP_{D}/\varepsilon+y$. The resulting function for large
$r$ is
\begin{align}\label{eq:p3asimpt}
\delta\Phi_3&=i\nu_3\left(\frac{P_{D}}{\varepsilon r}\right)^{(D-1)/2}
\exp\left(-\frac{\sqrt{8}P_{D}}{\varepsilon}\right)\\
&\times\left[i^{(D-1)/2}\exp(-i\sqrt{8}r)
-(-i)^{(D-1)/2}\exp(i\sqrt{8}r)\right] \ . \nonumber
\end{align}
The general solution of the left hand side of
\eqref{e:modes} can be written as a sum involving Bessel functions
$J_n$ and $Y_n$, which have the asymptotic behaviour
\begin{align}
& J_{\nu}(x)\to\sqrt{\frac{2}{\pi x}}\,\cos
\left(x-\frac{\nu\pi}{2}-\frac{\pi}{4}\right) \\
& Y_{\nu}(x)\to\sqrt{\frac{2}{\pi x}}\,\sin
\left(x-\frac{\nu\,\pi}{2}-\frac{\pi}{4}\right) \,,
\end{align}
for $x\to+\infty$. The solution satisfying the asymptotics given by
\eqref{eq:p3asimpt} is
\begin{equation}\label{eq:yasimpt}
\delta\Phi_3=\sqrt[4]{2}\sqrt{\pi}\,
\frac{\alpha_D^{}}{r^{D/2-1}}Y_{D/2-1}(\sqrt{8}r) \ ,
\end{equation}
where the amplitude at large $r$ is given by
\begin{equation}\label{eq:tampl}
\alpha_D^{}=2\nu_3\left(\frac{P_{D}}{\varepsilon r}\right)^{(D-1)/2}
\exp\left(-\frac{\sqrt{8}P_{D}}{\varepsilon}\right) \ .
\end{equation}
This is just the solution singular at $r=0$. The singularity is the
consequence of the initial assumption of exponential decay for large
$r$. The asymptotic decay induces an oscillation given by
\eqref{eq:yasimpt} in the intermediate core, and a singularity at the
center. In contrast, the QB solution has a regular center, but
contains a minimal amplitude standing wave tail
asymptotically. Considering the left hand side of \eqref{e:modes} as
an equation describing perturbation around the asymptotically decaying
solution, we just have to add a solution $\delta\Phi_3$ determined by the
amplitude \eqref{eq:tampl} with the opposite sign of \eqref{eq:yasimpt}
to cancel the oscillation and the singularity in the core. This way
one obtains the regular QB solution, whose minimal amplitude
standing wave tail is given as
\begin{equation}\label{eq:qbasimpt}\begin{split}
\phi_{QB}&=-\sqrt[4]{2}\sqrt{\pi}\,
\frac{\alpha_D^{}}{r^{D/2-1}}Y_{D/2-1}(\sqrt{8}r)\cos(3t) \\
&\approx -\frac{\alpha_D^{}}{r^{(D-1)/2}}
\sin\left[\sqrt{8}r-(D-1)\frac{\pi}{4}\right]\cos(3t) .
\end{split}\end{equation}
Adding the regular solution, where $Y$ is replaced by
$J$, would necessarily increase the asymptotic amplitude.

Subtracting the regular solution with a phase shift in time, we
cancel the incoming radiating component, and obtain the radiative tail
of the oscillon,
\begin{align}
\phi_{osc}&=-\sqrt[4]{2}\sqrt{\pi}\,
\frac{\alpha_D^{}}{r^{D/2-1}}\\
&\times\left[Y_{D/2-1}(\sqrt{8}r)\cos(3t)
-J_{D/2-1}(\sqrt{8}r)\sin(3t)\right]\nonumber\\
&\approx -\frac{\alpha_D^{}}{r^{(D-1)/2}}
\sin\left[\sqrt{8}r-(D-1)\frac{\pi}{4}-3t\right] .
\end{align}
The radiation law of the oscillon is easily obtained now,
\begin{equation}\label{e:symradlaw}
\frac{\mathrm{d} E}{\mathrm{d} t}=-3\sqrt{2}\pi^2
\frac{2\pi^{D/2}}{\Gamma\left(\frac{D}{2}\right)} K^2	
\left(\frac{P_{D}}{\varepsilon}\right)^{D-1}
\exp\left(-\frac{2\sqrt{8}P_{D}}{\varepsilon}\right)	.
\end{equation}

Without going into details in the case of asymmetric potentials, we
notice that once the one-dimensional problem is solved the first
dimensional correction of order $\varepsilon \ln \varepsilon$ is
obtained in essentially the same manner,
\begin{align}
K&=K^{D=1}\,\left(1-\varepsilon \ln \varepsilon\,i\,\sqrt3 C \right)
\,, \quad \lambda=\frac{5}{6}g_2^2-\frac{3}{4}g_3\,,\\
C&=\frac{i}{\lambda^2}\left(\frac{\lambda^2}{24}
-\frac{\lambda}{6} g_2^2+\frac{5}{8}g_5-\frac{7}{4}g_2g_4
+\frac{35}{27}g_2^2\right)\frac{8(D-1)}{15P_{D}}\,. \nonumber
\end{align}
The radiation law for small-amplitude oscillons is then
\begin{equation}\label{e:asymradlaw}
\frac{\mathrm{d} E}{\mathrm{d} t}=-\sqrt{3}\pi^2
\frac{2\pi^{D/2}}{\Gamma\left(\frac{D}{2}\right)} K^2
\left(\frac{P_{D}}{\varepsilon}\right)^{D-1}
\exp\left(-\frac{2\sqrt{3}P_{D}}{\varepsilon}\right).
\end{equation}

Eqs.\ \eqref{e:symradlaw} and \eqref{e:asymradlaw} can be interpreted
as an evolution equation for oscillons of the corresponding theory. As
the radiation power is exponentially small, the oscillons evolve
through undistorted quasibreather states adiabatically. The oscillon
energy was calculated in Ref. \cite{FFHL2} and is
$E=\varepsilon^{2-D}\,\frac{E_0}{2\lambda}+\varepsilon^{4-D}\,E_1$ as
a function of $\varepsilon$. Plugging this expression into the
radiation law gives us the evolution equation of $\varepsilon$.

\section{Numerical simulations}

In this section we will provide numerical results regarding oscillon
radiation. We have investigated a symmetric $\phi^6$ theory in $D=2$,
and the sine-Gordon theory in $D=2$ and $D=3$. The results of
numerical simulations in these theories confirm the theoretical
predictions with satisfactory accuracy. The time evolution of
oscillons was simulated using the fourth order method of line code
developed in Refs.\ \cite{FR,FR2}. Initial data was obtained by
applying the small-amplitude expansion method to $\varepsilon^3$
order. A one parameter fine tuning, by multiplying the amplitude with
a factor close to one, was applied. For $D=2$ this minimized the low
frequency modulation of the oscillon state, while in $D=3$ the tuning
was used to suppress the single decaying mode.

\subsection{Oscillons in the symmetric $\phi^6$ theory in $D=2$}

\begin{figure}[!htb]
\begin{center}
\includegraphics[width=8.5cm]{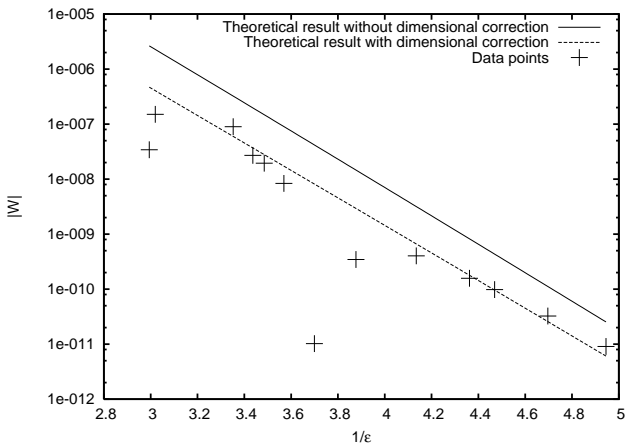}
\caption{Radiation law for small-amplitude oscillons in the $\phi^6$
  theory for $D=2$. We plotted the theoretical radiation law with the
  first dimensional correction and without it. \label{f:2phi6}}
\end{center}
\end{figure}

In Ref. \cite{FFHM} we have found that oscillons of the symmetric
$\phi^6$ theory defined by the potential
\begin{equation}
U(\phi)=\frac{1}{2}\,\phi^2-\frac{1}{4}\,\phi^4+\frac{1}{6}\,\phi^6\,,
\end{equation}
obeyed the theoretical radiation law in $D=1$ to a satisfactory
precision. In this theory the radiation of small-amplitude oscillons
is maximal among symmetric $\phi^6$ theories, therefore we have a
chance of observing the dimensional correction to the value of $K$.
The $\varepsilon$ dependence of the energy loss rate
$W=\mathrm{d}E/\mathrm{d}t$ is listed in Table \ref{table1}
and plotted on Fig.~\ref{f:2phi6}.
\begin{table}[!htbp]
\begin{center}
\begin{tabular}{|c|c|}
\hline
$\epsilon$  & $W$ \\
\hline
$0.33398$ &		$-3.4177\cdot 10^{-8}$\\
$0.33113$ & 	$-1.5053\cdot 10^{-7}$\\
$0.31125$ &		$-1.6574\cdot 10^{-7}$\\
$0.29829$ &		$-8.9890\cdot 10^{-8}$\\
$0.29098$ & 	$-2.7064\cdot 10^{-8}$\\
$0.28691$ & 	$-1.9475\cdot 10^{-8}$\\
$0.28017$ & 	$-8.3725\cdot 10^{-9}$\\
$0.27030$ & 	$-1.5860\cdot 10^{-11}$\\
$0.25794$ & 	$-3.4518\cdot 10^{-10}$\\
$0.24186$ & 	$-4.0292\cdot 10^{-10}$\\
$0.22925$ & 	$-1.5740\cdot 10^{-10}$\\
$0.22375$ &		$-9.8124\cdot 10^{-11}$\\
$0.21291$ &		$-3.2468\cdot 10^{-11}$\\
$0.20222$ & 	$-9.0850\cdot 10^{-12}$\\
\hline								
\end{tabular}
\end{center}
\caption{Radiation power $W$ for the oscillons in the $\phi^6$
theory. \label{table1}}
\end{table}
Analogously to the $D=1$ case, we find anomalous points, corresponding
to oscillons which radiate much slower than the theoretical
prediction. This reveals the rich structure of finite $\varepsilon$ corrections and
outlines the limitations of the calculation in the small-amplitude limit.

Based on the theoretical radiation law \eqref{e:symradlaw}, in the
$D=2$ case we use a semiempirical formula
\begin{equation}\label{e:empir}
\frac{\mathrm{d} E}{\mathrm{d} t}=
-K^2\frac{287.45}{\varepsilon}
  \exp\left(-\frac{6.18045\,b}{\varepsilon}\right),
\end{equation}
where an empirical correction constant $b$ is included in the
exponent, and the dimensional correction is included in the form of
\eqref{e:khidim}, with $i\sqrt{8}C=-1.592$ in this case. The
theoretical value of $K^{D=1}$ has been obtained in \cite{FFHM},
$K^{D=1}=0.579$, while the theoretical value of $b$ is obviously
$b=1$. We performed fits for the points with smaller $\varepsilon$
values than the last anomalous one. Fitting for both $K^{D=1}$ and $b$
for $\varepsilon<0.25$, we have obtained $K^{D=1}=0.05\pm0.02$ and
$b=0.82\pm0.02$. We get the theoretical pole term in the exponent quite
precisely, but the amplitude $K^{D=1}$ has large error. However,
setting $b=1$ and fitting only for $K^{D=1}$ in the same domain, we
have obtained $K^{D=1}=0.59\pm0.05$, which is close to the theoretical
prediction. Interestingly, there are points outside the domain of our
fits which also obey the theoretical radiation law. It is evident from
Fig.~\ref{f:2phi6} that the dimensional correction significantly
decreases the radiation power, but has little effect on the pole term,
i.e. the curves have approximately the same slope. The data points
back the theoretical form of the dimensional correction, even though
the precise $\varepsilon \ln \varepsilon$ functional dependence is
impossible to verify.

\subsection{Comments on the $\phi^4$ theory in $D=2$}

Oscillons in the $\phi^4$ theory show peculiar behavior in $D=2$. In
the $\varepsilon$ domain accessible to our lattice simulations they
obey the semiempirical radiation law, however it is not
consistent with the expected $\exp\left(-2\sqrt3 P_{2}/\varepsilon\right)$
pole term, even though the outgoing radiation is dominantly in the $\Phi_2$
mode. We remark that in the $\phi^4$ case the first dimensional
correction is so large in the accessible $\varepsilon$ domain that it
invalidates the computed leading-order behavior.

\subsection{Oscillons in the $D=2$ and $D=3$ sine-Gordon theory}

\begin{figure}[!htb]
\begin{center}
\includegraphics[width=8.5cm]{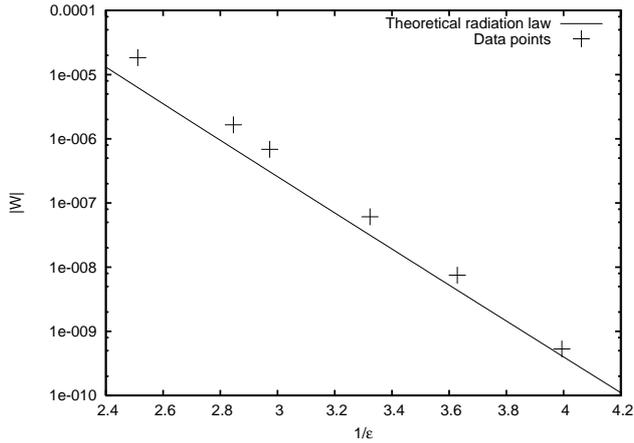}
\caption{The radiation law for oscillons in the $D=2$ SG theory.
\label{fig2}}
\end{center}
\end{figure}

\begin{figure}[!htb]
\begin{center}
\includegraphics[width=8.5cm]{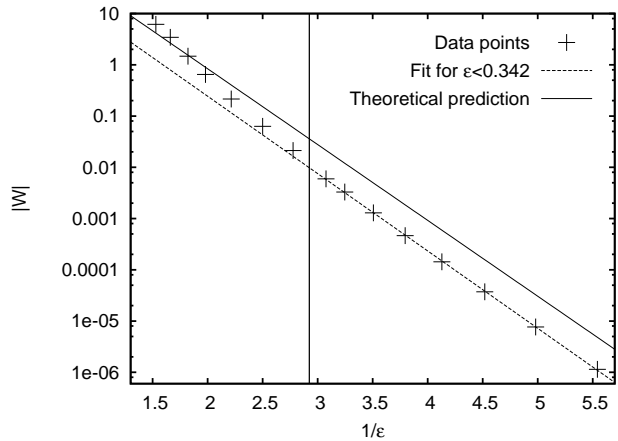}
\caption{The radiation law for oscillons in the $D=3$ SG theory. The
  vertical line corresponds to the energy minimum at
  $\varepsilon=0.342$. Points to the left of the line correspond to
  stable oscillon states, while the others to states having one decay
  mode, requiring a one-parameter tuning.  \label{fig3}}
\end{center}
\end{figure}

In the $D=2$ and $D=3$ SG theory we expect oscillons to live
exceptionally long, as the potential is symmetric and the dimension
independent part of $K$ is zero. Our
simulation results are collected in Table \ref{SG}, and plotted on
Figs.~\ref{fig2} and \ref{fig3}.
\begin{table}[!htb]
\begin{center}
\begin{tabular}[c]{|c|c||c|c|}
\hline
\multicolumn{2}{|c||}{D=2}&\multicolumn{2}{|c|}{D=3}\\
\hline
$\varepsilon$  & $W$&
$\varepsilon$  & $W$ \\
\hline
$0.39815$&$ -1.8395\cdot10^{-5}$&$0.32513$&$5.9630\cdot10^{-3}$\\
$0.35141$&$ -1.6531\cdot10^{-6}$&$0.30801$&$3.2942\cdot10^{-3}$\\
$0.33637$&$ -6.8685\cdot10^{-7}$&$0.28520$&$1.2931\cdot10^{-3}$\\
$0.30090$&$ -6.0986\cdot10^{-8}$&$0.26350$&$4.6372\cdot10^{-4}$\\
$0.27560$&$ -7.5049\cdot10^{-9}$&$0.24218$&$1.4336\cdot10^{-4}$\\
$0.25037$&$-5.3072\cdot10^{-10}$&$0.22130$&$3.7224\cdot10^{-5}$\\
 & & $0.20076$&$7.6601\cdot10^{-6}$\\
 & & $0.18042$&$1.1435\cdot10^{-6}$\\
\hline								
\end{tabular}
\end{center}
\caption{Radiation power in the SG theory.} \label{SG}
\end{table}

For the $D=2$ case, we can again use the semiempirical formula
\eqref{e:empir}, where now $K=\varepsilon M$. According to
\eqref{e:ksg}, the theoretical value of $M$ is $M=0.5502$, and for the
other parameter, obviously $b=1$. First we fitted both $M$ and $b$,
obtaining $M=1.79\pm0.16$ and $b=1.08\pm0.01$. Fitting only $M$ while
fixing $b=1$ we got $M=0.76\pm0.04$. Similarly to the $\phi^6$ theory,
the pole term can be measured quite precisely, but the amplitude has
larger errors.

Motivated by \eqref{e:symradlaw}, in the $D=3$ case we use the
semiempirical formula
\begin{equation}
\frac{\mathrm{d} E}{\mathrm{d} t}=
-K^2\frac{190.81}{\varepsilon^2}
  \exp\left(-\frac{3.40644\,b}{\varepsilon}\right),
\end{equation}
where $K=\varepsilon M$. The theoretical values are $M=1.9964$ and
$b=1$. The fit for $\varepsilon<0.342$ yields $M=1.156\pm 0.04$ and
$b=1.02047\pm 0.004$. Again, the pole term is precise, but the
amplitude has larger error. Setting $b=1$ and fitting only for $M$
does not help in this case, it gives about half of the theoretical
value, $M=1.0019\pm0.01$. Most probably this 
discrepancy is due the nonlinear effects related to the still too large values of
$\varepsilon$. Since the energy loss rate falls exponentially with
$\varepsilon$, our time evolution code cannot provide reliable results
for smaller $\varepsilon$ values.

\section{Acknowledgments}

This research has been supported by OTKA Grants No. K61636,
NI68228.

\end{document}